\documentclass[runningheads]{llncs}

\usepackage{graphicx}
\usepackage{amssymb}
\usepackage{epstopdf}
\usepackage[caption=false]{subfig}
\usepackage{csquotes}

\begin{document}
\title{A privacy-aware zero interaction smart mobility system}
%
%
\author{Stefano Righini\inst{1} \and
Luca Calderoni\inst{2} \and
Dario Maio\inst{2}}
\authorrunning{S. Righini et al.}
%
\institute{Quinck S.r.l., Italy
\email{stefano.righini@quinck.io}\\
\and
Department of Computer Science and Engineering, University of Bologna, Italy
\email{\{luca.calderoni,dario.maio\}@unibo.it}}

\maketitle              
\begin{abstract}
Smart cities often rely on technological innovation to improve citizens' safety and quality of life. This paper presents a novel smart mobility system that aims to facilitate people accessing public mobility while preserving their privacy. The system is based on a zero interaction approach whereby a person can use public transport services without any need to perform explicit actions. Operations related to ticket purchase and validation have been fully automated. The system is also designed with the privacy-by-design paradigm in mind, to preserve user privacy as much as possible. Throughout the paper several technical details are discussed as well to describe a prototype version of the system that was implemented. The prototype has been successfully tested in the city of Imola (Emilia Romagna, Italy) in order to prove the system validity on the field.

\end{abstract}
\section{Introduction}\label{sec:intro}

A smart city needs to face the effects of globalisation trends, processes of integration and urbanisation. These trends produce both benefits and drawbacks. Concerning drawbacks, addressing those related to urban mobility represents a considerable challenge. As an example, big cities are affected by air pollution, city crowding and so forth \cite{RGATE:Tomaszewska2018}.

Smart mobility belongs to the multidimensional concept of a smart city and pertains the aforementioned problems. Urban institutions are thus becoming increasingly interested in Intelligent Transport Systems (ITS) solutions which help them build a competitive city and enhance the citizens welfare \cite{DBLP:journals/itsm/SorianoSMGP18,DBLP:journals/giq/CledouEB18}.

Emerging technologies such as automated vehicles, peer-to-peer sharing applications and, more in general, Internet of Things (IoT), will revolutionise individual and collective mobility. This revolution will pose short and longer-term governance challenges \cite{RGATE:Docherty2017}.
As stated by \cite{RGATE:Batty2012} \enquote{new ICT is able to improve mobility on many levels thereby increasing spatial and aspatial accessibilities to jobs, leisure, social opportunities and so on, thereby enabling the citizenry to increase their levels of life satisfaction}.

Several definitions of Smart City were proposed across the years, but many of them agree on the fact that a Smart City, with the aim of improving the quality of life of citizens, requires multidisciplinary skills and massive use of the most disparate technologies, among which ICT often plays a decisive role. Integrated systems and services deployed in this direction include especially modern transport technologies \cite{RGATE:Mangiaracina2016}.

Through the years, many projects and research actions were devoted to applying intelligent solutions to urban mobility. The concept of Mobility as a Service (MaaS) emerged following a public-service-oriented point-of-view. This paradigm encapsulates the process of people utilizing a variety of private and public transport carriers without boundaries, where payment is tailored according to effective usage.
Some of them are also focused on reducing air pollution \cite{RGATE:Ydersbond2019}, while several other aim to control and manage the traffic flow from the municipality perspective \cite{DBLP:journals/eswa/CalderoniMR14}. Again, some works focus on driver support systems while other deal with decision support systems. Among them, infomobility and integrated mobility systems are probably those that better embody and implement the smart mobility and the MaaS concept, as they address both citizen and institution needs \cite{RGATE:Batty2012}.
Another key aspect concerning smart mobility is represented by its everlasting presence within smart city indicator frameworks. That is, smart mobility is (nearly) always considered among the aspects that contribute to raise city key performance indicators \cite{DBLP:conf/cata/DebnathPSRRMM20,DBLP:conf/ecsa/SanctisIRW20}.

In this paper we discuss a novel MaaS smart ticketing system designed with privacy in mind and with the aim of providing the user with a zero interaction experience. Adopting our system, the user is able to get on and off public transport avoiding any explicit action such as purchasing or validating a ticket. This combination represents a challenging aspect because, as evident from the literature analysis, when we push a smart mobility system in the interactionless direction, a privacy lack usually arises. 

The rest of the paper is organized as follows: in Section \ref{sec:related} we discuss the most relevant smart mobility systems with respect to the one we propose, and we underline why and to what extent our system differs with them. In this section we also discuss privacy implications concerning smart mobility; in Section \ref{sec:arch} we detail the technical aspects of our system; in Section \ref{sec:discussion} we discuss the most relevant features of the system we propose and we highlight the novel aspects concerning privacy-by-design and zero interaction.

\section{Related works}\label{sec:related}

Through the years, many smart mobility systems have been proposed in literature. While a comprehensive list of them falls out of the scope of this work, we deem important to highlight some systems that deserve attention due to its prominence or due to its similarities or differences with respect to the one we present in this paper.

Many researchers have devoted their efforts to traffic analysis, prediction and simulation. These systems usually rely on a three layer architecture (sensing, storage, service) where the second layer, also referred to as middleware, exposes several API to receive and provide the collected data \cite{DBLP:conf/smartcomp/NesiBBCMP16}. For instance, the SEnTINEL application was developed to provide a real-time tracking service \cite{DBLP:conf/dcoss/QuessadaCJLM19}. SEnTINEL consists of an Android-based mobile application which acquires user routes, and a web application that makes the acquired data available to various stakeholders. Possible SEnTINEL applications include vehicle routing services as well emergency routes discovery. Again, the $S^2$-Move project proposes an architecture able to collect, update, and process real-time and heterogeneous data  derived from a number of different sources (sensors, smartphones, vehicles, pedestrians) in order to provide innovative mobility services \cite{DBLP:conf/iscc/MarchettaNPSS15}.

Among those technologies that enable mobility data gathering, in recent years Bluetooth devices or Wi-Fi access points were extensively applied as a source of information for traffic data acquisition \cite{DBLP:journals/itsr/StevanovicOGGK15}. Bluetooth devices are unaffected by weather or light conditions and their deployment has a low economical cost. As an example, the city of Valencia has developed a Bluetooth-based project aimed at analyzing mobility in the urban area\footnote{https://www.uraia.org/en/library/inspiring-practices-catalogue/valencia-smart-city-platform/}. Signal poles in streets featuring high traffic density are coupled with Bluetooth devices, allowing the detection of vehicles in different areas when they approach the sensors.

Similarly, \cite{DBLP:journals/ijcssa/Fernandez-AresG16} propose a monitoring system based on the detection of Bluetooth and Wi-Fi signals gathered from mobile and portable devices. The system is designed to analyse urban traffic through several detection nodes. It provides automatic and accurate traffic monitoring, and it is effectively applied to synchronise traffic lights.

While the aforementioned systems are mostly devoted to traffic monitoring, in this paper we rely on the Bluetooth technology to design a different type of ITS architecture.

The solution we discuss in this paper falls under the smart ticketing and infomobilty umbrella. The fusion of urban ticketing systems and new technologies paved the way for the MaaS paradigm, where citizens are able to move using a combination of different transport carriers (self-driving cars, car sharing, subway, buses etc.) while relying on a single integrated system for reservation, payments and so forth.
A holistic approach to shared mobility and multi-modal mobility in ITS is discussed by \cite{DBLP:conf/icsoc/KusterMS17}. The authors provide a roadmap towards an agent-based interactive mobility assistant, while they do not provide an effective implementation or case study.
\cite{DBLP:conf/qrs/KarinsaloH18} discuss a novel system which utilizes QR code that stores and uses travel information in smart contract over Ethereum. In this MaaS ecosystem all travel details are stored in a single ticket. The user needs to print the QR code, implying an explicit interaction with the system. The authors do not face privacy issues nor they provide a zero interaction design as we propose in our solution.

Concerning ticketing systems, one of the main issues that inhibits the zero interaction principle is represented by the travel planning/reservation or by the payment procedure. These steps usually require the user to perform an explicit action on a mobile application or on a dedicated website. This is the case of the Seattle online ticketing system, that provides online payment of parking and traffic fines, or the Barcelona WeSmartPark system\footnote{https://barcelona.wesmartpark.com/ca}, providing parking availability, booking and payment.

One of the systems that mostly fit the one we discuss in this paper is \emph{OV-chipkaart}\footnote{https://www.ov-chipkaart.nl/}, the integrated mobility system deployed in the Netherlands. The system relies on contactless smart cards and was firstly introduced during April 2005 in the Rotterdam subway. Subsequently, it was extended to other municipalities and to other transport carriers such as buses and railways. From 2014 onward \emph{OV-chipkaart} became the sole ticketing system adopted in the Netherlands. At the start of each journey, the citizen holds the contactless card against the screen on a gate or card reader. If the credit is sufficient the gate opens or the card reader beeps to confirm and a green light appears. At the end of the journey, the smart card  needs to be placed against the gate or card reader again to check out. The display shows the cost of the journey and the remaining credit. Currently, three types of cards are supported: fire and forget, anonymous and personal. The first type does not differ as much from a traditional ticket. The second one is a disposable card with a fixed amount of credit, while the third is a rechargeable debit card. One of the most relevant issues when the system was deployed was related to users privacy. Privacy was not considered at all during the system design, and whoever spotted the Id number from the card could be able to check the movements of the card holder on the system website. While this privacy breach was addressed in the past years, the system does not comply with the zero interaction principle. Each time the citizen uses a transport carrier he/she needs to explicitly tap on readers two times (at the beginning and at the end of the ride). 

Another system that deserves to be discussed is \emph{Fairtiq}\footnote{https://fairtiq.com}, an integrated mobility system designed in Switzerland. It provides the user with a mobile application that enables smart ticketing in Switzerland, Liechtenstein and in several other municipalities. Several transport carriers are supported, such as railways, buses and boats. The user needs to tell the application that he is starting (or ending) his journey and the mobile app does the rest: relying on the GPS sensor and on other smartphone related sensors, the system derives which kind of carriers the citizen used and computes the corresponding fees. While this system is designed with the zero interaction principle in mind, it does not offer any privacy-by-design guarantee. Differently from the system we propose, \emph{Fairtiq} constantly tracks the user movement through GPS and collects an impressive amount of personal data.

\subsection{Privacy issues in modern mobility systems}\label{sub:related-privacy}

Smart mobility systems often rely on a number of sensing technologies and ICT infrastructures in general. This condition raises several privacy concerns.

Privacy issues play a key role within the smart city context, where citizens are subject to an ubiquitous and constant interaction with several ICT infrastructures and systems \cite{DBLP:journals/comsur/EckhoffW18}. With the widespread adoption of GPS-based devices, a massive amount of location data is being collected. As smart mobility systems often rely on the smartphone GPS signal, citizens that use these kind of systems risk to be constantly tracked. Location traces represent sensitive data and may reveal an impressive number of habits the user would not intend to share \cite{DBLP:journals/comsur/PrimaultBMB19}. The interest for location data has sensibly increased as they underlie strategic information, enabling a wide range of location based services \cite{DBLP:journals/tmc/ChonTSC14} and allowing to perform a very precise and invasive user profiling.

Unfortunately, a wide number of studies proved that location trajectories may be de-anonymized to discover users identity \cite{DBLP:journals/jcss/GambsKC14,DBLP:journals/tissec/JiLSHB16} or to learn their relevant stay points \cite{DBLP:journals/istr/FranciaGGS20}.

During the past years it was proved that users' privacy may be threatened even when the user adopts privacy-friendly naive solutions, as anonymous multiple-ride tickets \cite{DBLP:journals/ijinfoman/AvoineCDM014}.

Privacy concerns related to smart mobility systems and, more specifically, to IoT-based mobility systems \cite{DBLP:conf/eucnc/GebruCCG20} were further increased by the growing trend of deploying these solution on the Cloud. Despite the loss of independence it implies, the adoption of cloud-based commercial middlewares represents the most common solution to deploy an IoT-based system. In addition, the majority of open source IoT frameworks are designed with cloud platforms in mind \cite{DBLP:journals/jsa/CalderoniMM19,DBLP:conf/wf-iot/CalderoniMM19} and are usually deployed on Infrastructure as a Service (IaaS) providers. From the one hand, cloud platforms are subject to several issues concerning communication channel security or context security \cite{DBLP:conf/IEEEares/Calderoni19}; from the other hand, data outsourcing pose a serious threat to users' privacy.

Unfortunately, it is usually very hard to quantify the amount of privacy offered by a digital service. When researchers and professionals deal with privacy issues of a system, they often introduce ad hoc privacy metrics instead of using existing ones, making their claims incomparable. To this end, Wagner and Eckhoff proposed a comprehensive taxonomy of privacy metrics which we adopt throughout this paper \cite{DBLP:journals/csur/WagnerE18}. Following this principle, we refer to one of the most used privacy metrics, namely $k$-anonimity \cite{DBLP:journals/ijufks/Sweene02}, and we also rely on data \emph{anonymization} and \emph{generalization} \cite{0829/14/ENWP216}, standard techniques adopted by the EU General Data Protection Regulation (GDPR).

\section{System architecture}\label{sec:arch}
The main focus of our system is to offer users an intuitive and fully automated way of ticketing. Considering the current problems and challenges of the transport systems the solution that we propose aims to improve the usability while taking care of users' privacy. The solution allows users to use public transport services in a completely transparent manner, i.e, without performing any action concerning purchase and validation of tickets. Furthermore, the ultimate goal of the system is to provide users with a unique platform to access different public transport services, such as buses or subways.

\begin{figure}[t!]
    \centering
    \includegraphics[scale=0.45]{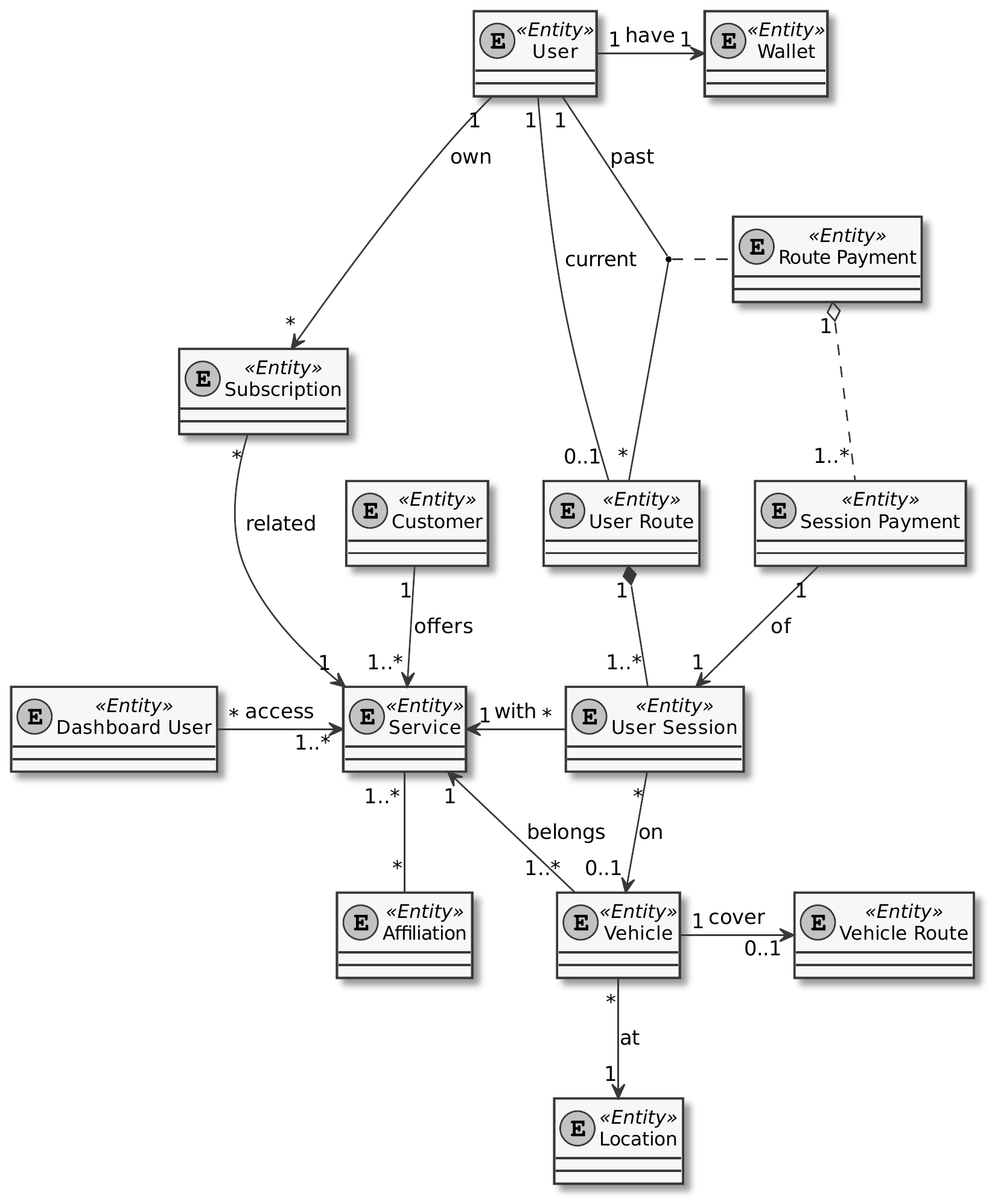}
    \caption{Entity model of the system using UML 2 class diagram.}
    \label{fig:entity_model}
\end{figure}

Figure \ref{fig:entity_model} shows a conceptual schema of the main entities and realtionships of the proposed system.

\textbf{User Route} A User Route is a representation of a route taken by a user. A route could be considered as a journey where, given a start and end point, several different vehicles and transport services can be used to complete it.

\textbf{User Session} A trip made by a user with a public transport service is defined as a User Session. Every time a certain user travels with a certain transport service an instance of this entity is created. Within a User Route there may be one or more User Sessions.

\textbf{User} Each entity of this type identifies a user within the system and corresponds to a single natural person. Two different users corresponding to the same natural person cannot exist within the system and vice versa.

\textbf{Service} Identifies a particular service offered by a transport agency and usable through the system. Each company may have one or more services and for each of these an entity of this type is generated.

\textbf{Wallet} These are the users' virtual wallets and contain all the information about the amount of money in the wallet, the payment methods connected, the configuration for money transfer.

\textbf{Route Payment} Once a route has been completed, the price that the user would have to pay is computed, considering the costs of the sessions within and any discounts that may be applicable. For example, any affiliation or contract among the customers and a transportation provider as well as any subscription a user may hold for a given service are taken into account.

\textbf{Session Payment} It refers to the cost of a single session and is calculated according to the prices and terms defined by the customer offering the service used. In this sense, only the single session is considered without taking into account any other sessions within a route. The real cost to be applied to the user will be computed at the end of the route.

\textbf{Dashboard User} Each transport company can create several profiles in order to access the dashboard and view the data produced by the uses for the services it offers. This entity represents a user who can authenticate himself in the dashboard to view information about one or more of the services of the agency he works for.

\begin{figure*}[t!]
    \centering
    \includegraphics[scale=0.6]{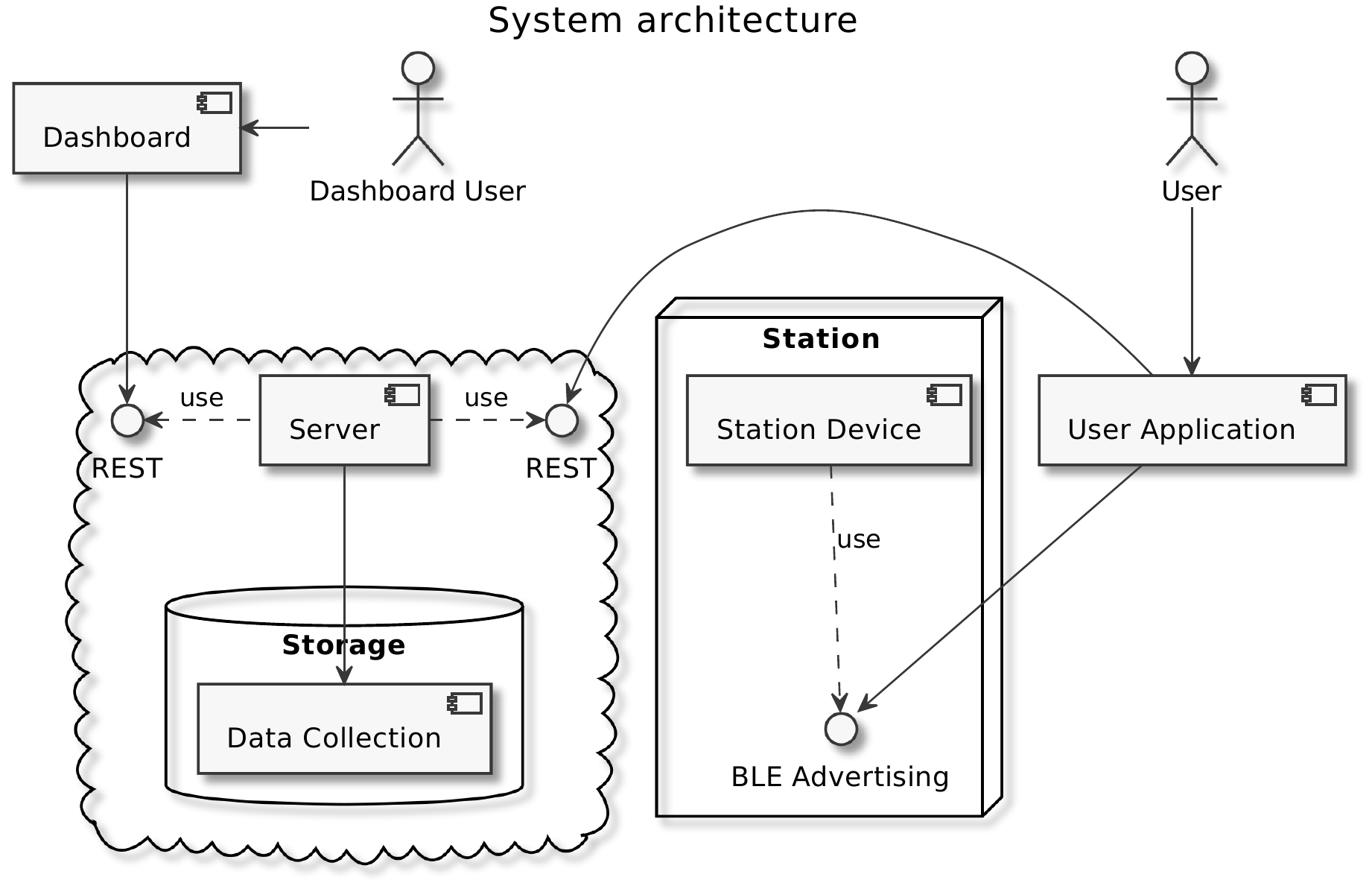}
    \caption{System architecture using UML 2 component diagram.}
    \label{fig:system_architecture}
\end{figure*}

The system consists of four components: a mobile \emph{User Application}; a \emph{Station}, to be installed on a vehicle or on a turnstile; a \emph{Server}, designed with the Privacy-by-Design principle in mind and relying on a microservices architecture; a \emph{Dashboard}, consisting in a web application allowing transport agencies and institutions to visualize the gathered data. In the following sections we detail the Server and the Station because of the important role they play in the system. The protocol we adopted to implement those goals and novel features is also discussed. The architecture of the system is shown in Figure \ref{fig:system_architecture}.

The server side components, Server and Storage, are deployed in the cloud. Specifically, the Server let the User Application and Dashboard instances communicate one each other through a REST interface over the HTTPS protocol.

The offered transport services are divided in two categories depending on the access mode. On the one hand we have those where access is provided by means of a turnstile, while on the other hand we consider those that require the user to validate a ticket on board.

A Station could be installed inside a vehicle (i.e. in the case of a bus) or in a turnstile (i.e. in the case of a subway). Each one of them exchanges information using the advertising mode of the Bluetooth Low Energy (BLE) protocol. A User Application in proximity of a Station can read the shared data.

\subsection{Station}\label{sub:station}
A Station is an IoT device to be installed inside vehicles, designed for buses, trains, or turnstiles. This device has both hardware and software components. The first one needs a System on Chip (SoC) equipped with a Global Positioning System (GPS) sensor and a BLE communication interface. Moreover a station installed on a turnstile needs a network interface in order to ask the server the permission to open it.

On the other side the software part has been designed with an Event Loop and Event Driven model. This choice allow the station to handle multiple tasks and to have the same software architecture model for both station types. The one installed on turnstiles only adds the behaviour concerning authorization and opening of the turnstile.

Summarizing, the Stations continuously share their identifier and location, updating the latter periodically. In addition, the stations installed in the turnstiles authorise their own opening through a remote request.

In order to test the system and prove its feasibility, a prototype of a Station has been realized. The adopted hardware consists of a RaspberryPi 3 Model B+, equipped with a GPS sensor. Concerning Bluetooth, it is natively installed on this board (version 4.2/BLE). To retrieve the station coordinates, the GY-NEO6MV2 NEO-6M sensor was adopted.

\subsection{Server}\label{sub:server}
The Server has been designed adopting a microservice architecture, so as to isolate each single component in a dedicated microservice and to enable horizontal scalability.

\begin{figure}[htbp]
    \centering
    \includegraphics[scale=0.42]{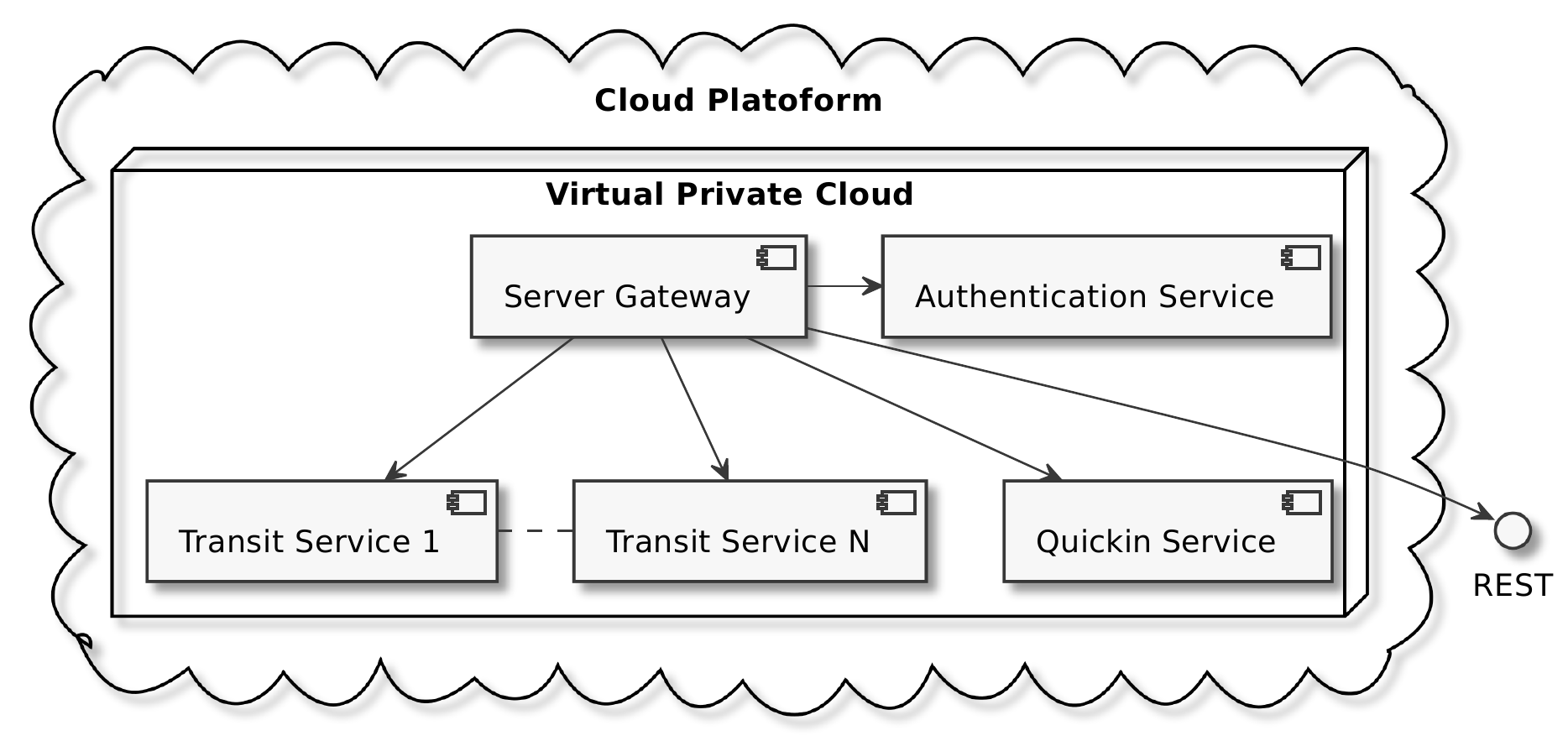}
    \caption{Architecture of the Server expressed using UML 2 component diagram.}
    \label{fig:server_architecture}
\end{figure}

Microservices can be divided into three groups: the first one, consisting of \emph{Server Gateway} and \emph{Authentication Service}, exposes the system features through a REST interface and takes care of authentication; the second one includes the general services of the system, common to all the transport services; the third one, finally, takes care of managing the specific transport services. These components are shown in the Figure \ref{fig:server_architecture}, where the Transit Service and the Quickin Service represent the third and the second group respectively.

The Server Gateway is the only component that can be reached from the outside. For this reason, it exposes all the functionality made available by the system to Dashboards and User Applications.

Quickin Service is the block element responsible for providing the functionality of user and customer information. In fact, none of the components is designed to handle remote requests directly. Similarly to the Server Gateway, Quickin Service exposes all the functionalities that are made available by each other component through a dedicated interface. It is implemented with its own internal logic so that it is able to offer complex functionalities composed of several operations, involving different microservices.

Similarly to the Quickin Service, the Transit Service comprises a set of microservices dedicated to a specific transport service. Thus, for each different transport service we deploy a complete block composed of each micoservice belonging to the the Transit Service. A basic implementation of this block is provided relying on the type of transport service (bus, train, metro, etc.). Further customization can be applied as well, depending on specific needs that follow a company-based perspective.

Using a different set of microservices for each different transport service leads to a complete separation of the information of each trip. This approach also induce a higher level of scalability. Furthermore, customization can be applied to a specific group of microservices without affecting any other group.

It is important to point out that groups decoupling also facilitates to separate the information into several different databases. As an example, data concerning the users rides (User Session) are isolated and stored separately. Thus, they can be disclosed following a secure and controlled procedure to those partners and customers asking for public transport statistics. In addition, these data are anonymized through generalisation and $k$-anonymity techniques before being stored in the respective database (as further discussed in Section \ref{subsub:datamodel}).

\subsubsection{Microservices}\label{subsub:microservices}

\begin{figure*}[t!]
    \centering
    \includegraphics[width=\textwidth]{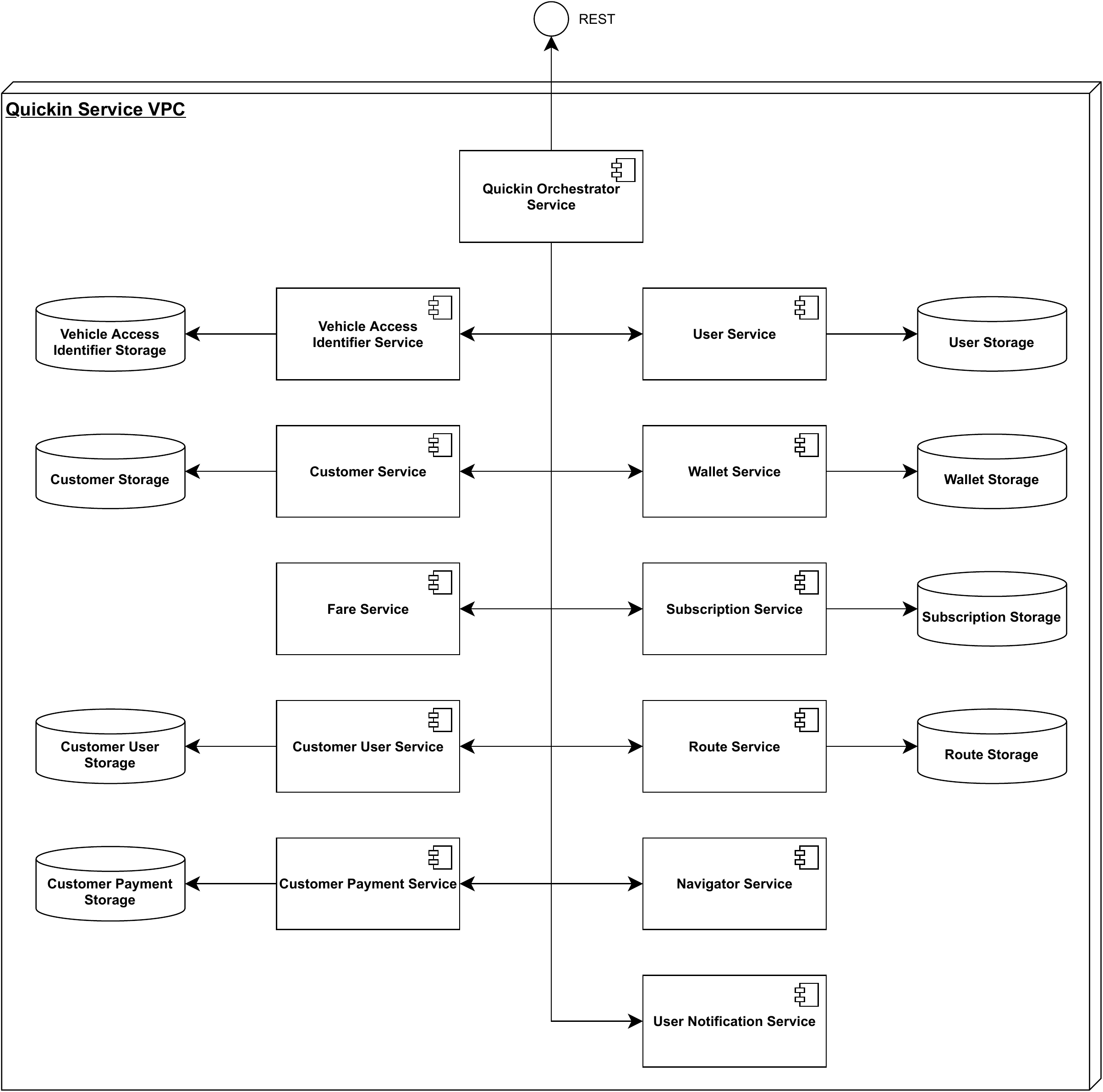}
    \caption{UML 2 component diagram representing the Quickin Service architecture expanded.}
    \label{fig:quickin_service_architecture}
\end{figure*}

\begin{figure*}[t!]
    \centering
    \includegraphics[width=\textwidth]{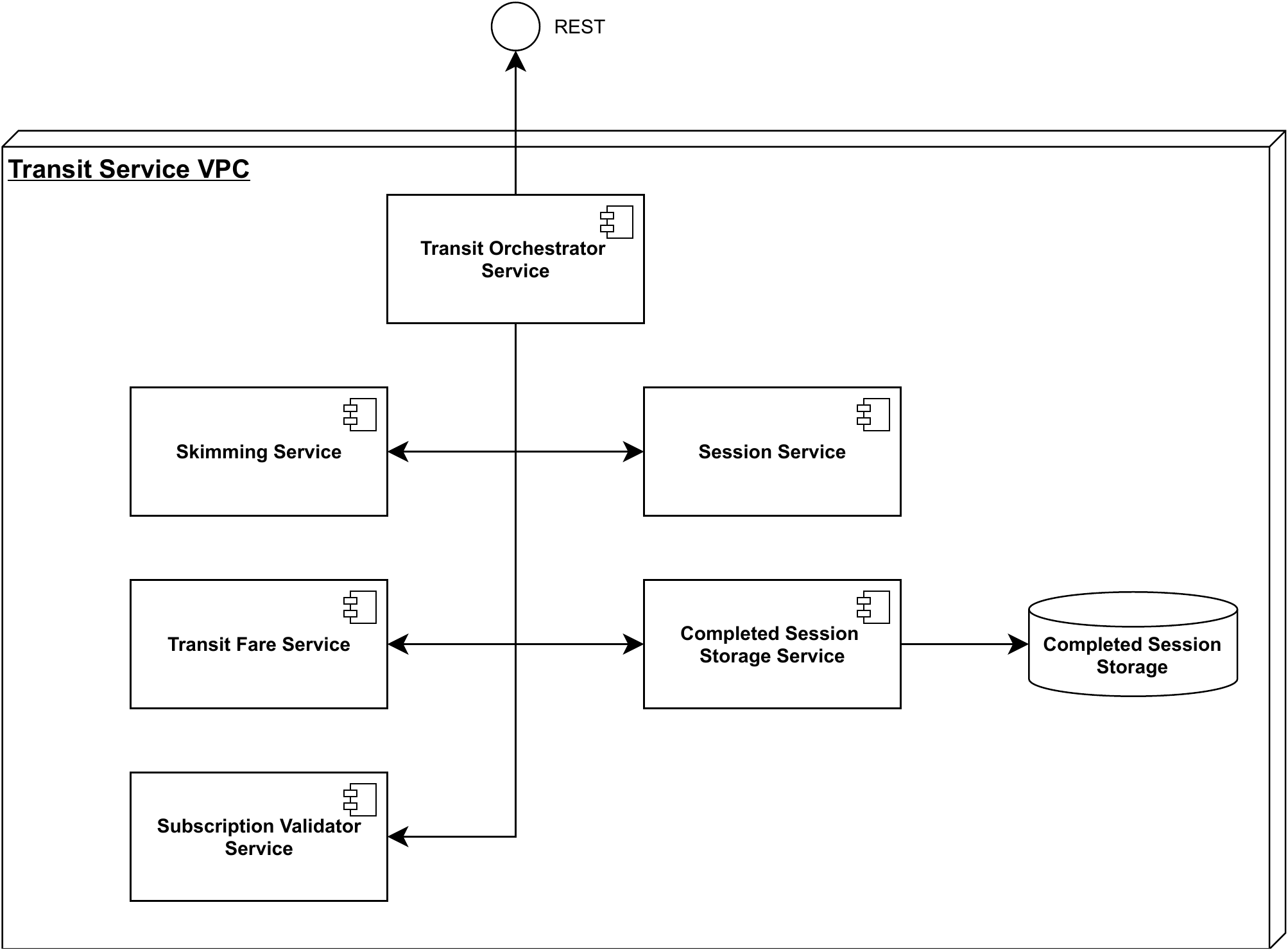}
    \caption{UML 2 component diagram representing the Transit Service architecture expanded.}
    \label{fig:transit_service_architecture}
\end{figure*}

A comprehensive representation of the microservices composing the Quickin Service is provided in Figure \ref{fig:quickin_service_architecture}. Similarly, Figure \ref{fig:transit_service_architecture} shows the microservices that are involved in Transport Service. The most relevant microservices are discussed below.

\paragraph{Quickin Service}

The User Service and the Customer User Service are respectively responsible for user management. Specifically, the first module handles end-users while the second one handles partners accounts designed to interact with the dashboard.

The Wallet Service manages user's virtual wallet, payment methods and all those operations involved in wallets charging in relation with user routes.

The Customer Payment Service is responsible for money transfers to partner transport companies.

The Customer Service is in charge of managing all the transport services included in the system and the information related to them.

The system provides the user with a navigation feature through the Navigator Service.

The Route Service manages the routes (User Route) taken by users. A route consists of the usage of one or more transport services within a user's journey. Therefore, each route may consist of several sessions (User Session).

Finally, the Vehicle Access Identifier Service handles the associations between a Station, the vehicle or turnstile in which it is located and the service it belongs to. This way, if a station were to be reallocated to another service, it would be sufficient to update this association accordingly, avoiding any direct maintenance on the station software.

\paragraph{Transit Service}

As mentioned above, for each transport service offered by a separate company, a specific set of microservices is deployed. In turn, when a partner company possesses different transport carriers, a complete set of dedicated microservices will be deployed for each of them.

The Data Storage Service, in fact, only stores the completed sessions for the transport service it deals with.

The Session Service deals with ongoing sessions management, and it mostly collects data used for validating the session itself, i.e., to determine which service is being used by the user. It should be noted that this microservice is not coupled with any database as ongoing sessions are stored in RAM. This approach is motivated by the wide amount of data that are recorded during each session. A DBMS-based approach would lead to a large number of write operations and would significantly affect performances.

Relying on the data collected by the Session Service, the Skimming Service determines the final session validation once it has been concluded.

Finally, the Transit Fare Service is responsible for determining the cost of a single session according to the specific plan offered by the partner company.

\subsubsection{Data model}\label{subsub:datamodel}

In order to implement a distributed architecture and to isolate the microservices, a separate database was deployed for each relevant service. Concerning the DBMS, it should be considered that the system deals with different transport services, different clients and evolving regulations. In addition, the number of involved partners and end-uses may increase over time, resulting in an unpredictable data increment. Consequently, it would be advisable to rely on a scalable database engine. Again, since data are not subject to a predefined model, a flexible data storage would be preferable to allow future changes.

For the aforementioned reasons, a NoSQL architecture was adopted. Specifically, we relied on the DynamoDB DBMS offered by Amazon Web Services (AWS), which provides several integrated features to implement an efficient horizontal scalability strategy. The frequent modification of the stored data objects is supported as well. DynamoDB is designed to handle large amounts of data and preserves a latency in the order of milliseconds independently of the request peaks.

As discussed above, rides made by users are stored in different databases. The Completed Session Storage database (see Figure \ref{fig:transit_service_architecture} for reference) servers as example to to discuss the data model behind a user session. To accomplish the EU privacy regulation, a privacy-by-design approach (see Section \ref{sub:privbydesign} for more details) has been adopted and the user sessions have been anonymized through generalisation and $k$-anonymity techniques. Table \ref{tab:completed_user_session_data} details the introduced data model. 

\begin{table}[ht]
\centering
    \begin{tabular}{|l|l|}
    \hline
    Field & Description \\
    \hline
    Gender & The gender of the user \\
    Age range & The age range of a user \\
    Starting position & Latitude and longitude\\
    Arrival position & Latitude and longitude\\
    Starting timestamp & Date and time \\
    Arrival timestamp & Date and time \\
    \hline
    \end{tabular}
    \caption{Information saved for completed User Session saved in the Completed Session Storage.}
    \label{tab:completed_user_session_data}
\end{table}

Furthermore, to strengthen privacy protection and the system security, all data are subject to encryption before being stored. Decryption occurs only when necessary and is performed by the system itself. The databases will therefore only contain encrypted information.

The underlying encryption protocol is referred to as \emph{envelope encryption} and is offered by AWS as a DynamoDB integrated feature. Since key management as well as encryption and decryption procedures take place on the provider side, AWS is to be intended as a trusted party in our privacy and security model.

As the number of stored information is limited to the essential, the system does not require high performance in accessing and updating information. Thus, cryptographic operations do not impact as much on the system functionality.

\subsection{Communication protocol}\label{sub:communication_protocol}

\begin{figure*}[t!]
    \centering
    \includegraphics[scale=0.47]{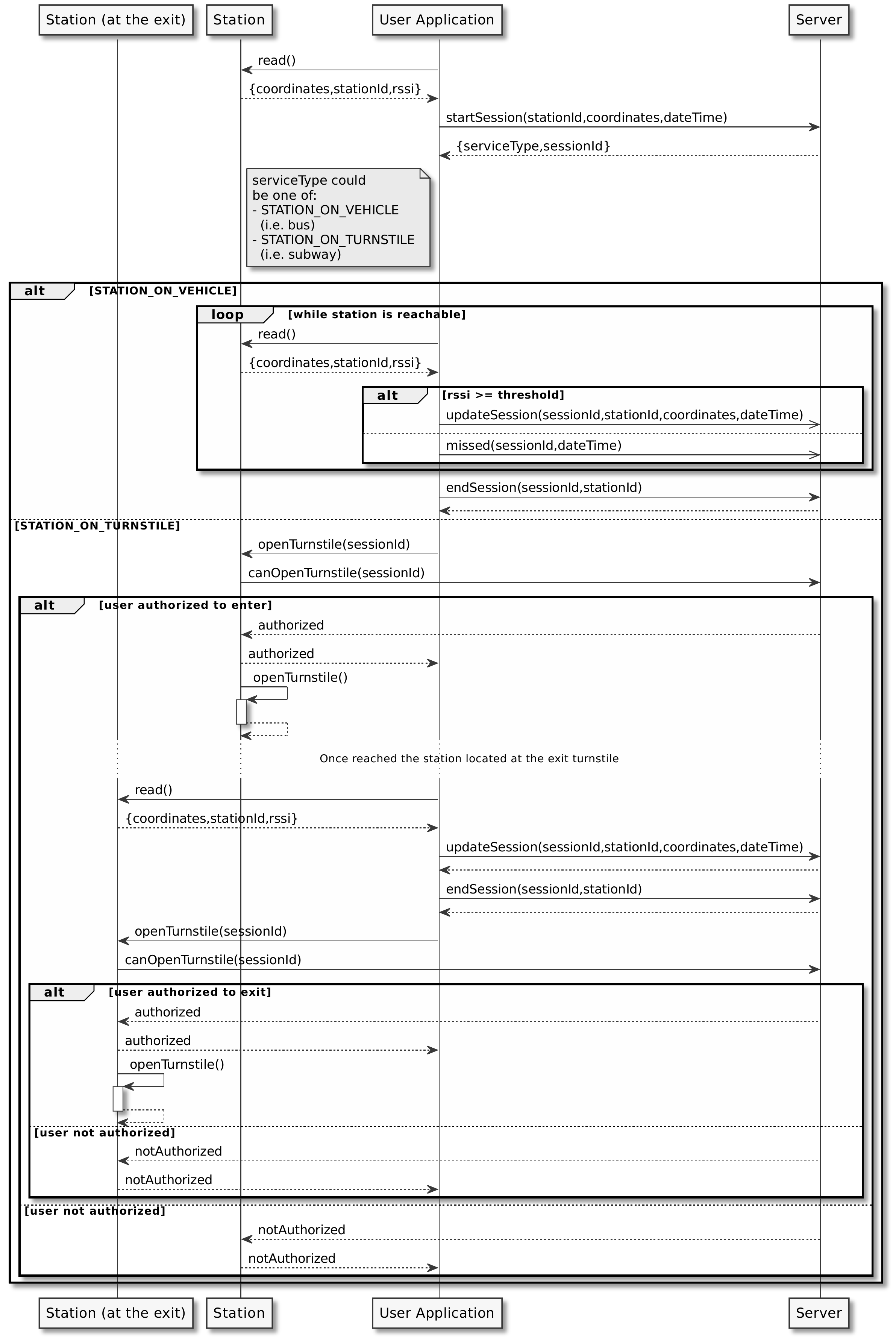}
    \caption{UML 2 sequence diagram representing the interaction between Station, User Application and Server for a single User Session.}
    \label{fig:single_session_sequence_diagram}
\end{figure*}

As formerly discussed, a User Route is a set of one or more sessions. In order to better understand our ITS solution it is important to clarify how the second one has been implemented in the system.

The protocol designed to handle a User Session is described in Figure \ref{fig:single_session_sequence_diagram}. This diagram helps deducing the zero interaction property of the system, as each communication step is performed without the need of human actions.

Before discussing the protocol details we introduce an exemplary use case reflecting a user interaction with a transport service. When the user approaches the system for the first time, he must install the dedicated mobile application and proceed with registration and authentication. Once these operations have been completed, he can close the application, which runs in background, and let the smartphone lay wherever he wants. The interaction with a transport carrier may be divided into four steps:
\begin{enumerate}
    \item The user boards on a participating vehicle (i.e. a vehicle with a station installed within); the mobile application autonomously starts communicating with the station.
    \item When the user reaches his/her destination, he/she gets off the vehicle. The communication stops and the end of the trip is recorded (User Session).
    \item The collected data are analysed in the cloud to determine whether or not the ride has effectively taken place. Information is stored in an dedicated database for future analysis.
    \item If the validity of the ride is confirmed, the system automatically charge the user's wallet.
\end{enumerate}

We now proceed to the technical details of the protocol depicted in Figure \ref{fig:single_session_sequence_diagram}.

When a User Application and a Station are close enough (and the signal strength is above a certain threshold) the former reads the information shared by the latter through the BLE Advertising protocol. The User Application sends a request to the Server to communicate the start of a new User Session, and receives a session identifier as response. The type of service it is connected to is included in the response as well.

From this point onward, the protocol differs according to the type of service, namely \emph{Station on board} or \emph{Station at the turnstiles}.

In the first case the User Application communicates periodically with the Station until it is reachable. At each step, if the signal strength expressed in RSSI is higher than a certain threshold, the application sends the Server the new data, updating the current User Session. Otherwise, it informs the server that the information is missing. Missing data are important as well since they are further analyzed by the Server to validate a User Session. At the end of the run, when the Station is no longer reachable, the User Application notifies the Server.

In the second scenario, the User Application requests the Station to open the turnstile. This request is propagated in turn to the Server which checks user permissions. If the user is authorized and meets the requirements to start a User Session, the Station opens the turnstile and the session begins. The user then travels to his destination where he finds a Station controlling the exit turnstile. The User Application reads the information from the Station, sends it to the Server for update, and immediately communicates the end of the session. Finally, following the same procedure applied at the entrance, the User Application requests the Station to open the turnstile to allow the user to exit.


\section{Discussion}\label{sec:discussion}
In this section we discuss the most innovative aspects of the solution we implemented and we motivate their importance. First, we highlight the potential of a zero interaction approach within the smart city domain. Second, we illustrate how we deal with privacy concerns through an analysis of the privacy enhancing technologies we adopted. 

\subsection{Zero interaction}\label{sub:zeroint}
As explained above, one of the strengths of our system is that it speeds up both ticket purchase and validation through a completely automated and transparent procedure from the user perspective. This achievement is reached relying on IoT technologies such as BLE.

As mentioned in Section \ref{sec:related}, \emph{OV-chipkaart} is one of the most prominent smart mobility systems operating within the EU, currently adopted in the Netherlands. Differently from our system, \emph{OV-chipkaart} allows access to public transport services through a rechargeable smart card that needs to be explicitly validated at the beginning and at the end of each journey. Moreover, it is required for the user to periodically recharge his card at specific recharging stations or at the ticket office.

The solution proposed in this article is based on a real Business Process Reengineering (BPR) of the purchase and validation events. In fact, after the first registration, the user will no longer be required to perform any action relating to these operations.
In addition, the ticket payment process has also been automated. At the end of each user route, the system automatically computes the amount due and charges the wallet using the preferred payment method set by the user.

In the smart city context, increasing the use of public and shared transport services is very important. This condition enhances city life from several perspectives. For example, it helps reducing both air pollution and road congestion. One of the main factors that makes public transport attractive is the way to access it. The simplification and optimisation achieved with the zero interaction mode proposed in this article represent a significant incentive for the end user to choose this form of mobility. Several reasons supporting this thesis are discussed below.

One of the major advantages of this approach is time saving, as a user no longer has to spend time choosing, buying, and validating a ticket. The concept of time saving is at the heart of the development of new smart cities. Several articles give examples of how technological progress can optimise processes in order to reduce the amount of time spent by people in performing city-related tasks. As an example, a study showed that $35\%$ of the total traffic in smart cities is due to people looking for a parking slot \cite{341072337_Building_Smart_Cities_Applications_based_on_IoT_Technologies_A_Review}.

Assuming a person uses the bus once a day and takes five minutes to perform the aforementioned actions, he or she would save $35$ minutes in a week. In addition, if a citizen arrives at the train station just in time for departure, he could immediately board the train without worrying about the ticket. These are just two examples that highlight how important time is nowadays -- specifically in the smart city context -- and how this system can help people in this direction.

Again, removing the need for ticket-related actions, the system drastically reduces the number of people queuing to buy a ticket (at ticket machines or ticket offices) and thus improves the purchasing process. The same applies to validation at ticket machines. The absence of active validation can neutralise the crowding phenomenon near ticket machines. Avoiding crowding in public transport is particularly advisable during a pandemic as the Covid-19 one, as it could help preventing the spread of infection.

Another key aspect for smart cities is to ensure (as far as possible) the same opportunities for all citizens, including those with disabilities. The proposed system can certainly fit this condition, since no interaction is required to validate tickets nor to buying them. The sole action to be performed is making the journey.

Finally, even elderly may benefit from the zero interaction feature. Although they may be unfamiliar with digital systems, this technology could simplify their journey by allowing them to use public transport relying on a completely automated process.


\subsection{Privacy by design}\label{sub:privbydesign}
As underlined since the paper introduction, we designed this system with privacy in mind. Users privacy indeed represents an increasingly relevant problem, especially when personal data are transferred to third parties for further -- unpredictable -- analysis.

First, we decided to strengthen the system security forcing encryption on all data intended for storage. To this end, we set up envelope encryption in the provider back-end. While this technique increases the privacy level in case of data theft deriving from hosting platform breaches, it does not prevent the cloud provider (AWS) to decrypt the content. Thus, the hosting provider is to be considered a trusted party.

A more relevant aspect concerning privacy is that our system does not rely on the smartphone GPS sensor. Localization data are indeed derived from the Stations. Moreover, the mobile application designed to operate in combination with the Stations does not steal any other user information and does not collect any other information derived from mobile sensors (such as magnetometer, accelerometer etc.). This behavior is very different from the one adopted by \emph{Fairtiq}, a notable mobility system operating in central Europe. \emph{Fairtiq} relies on user GPS and continuously collects its signal. Differently from the system we propose, \emph{Fairtiq} constantly tracks the user movement through GPS and collects an impressive amount of personal data.

Finally, to further enhance privacy protection, we decided to separate user data through isolated microservices. Personal data are stored in a different location with respect to sessions data. Moreover, as sessions data may be transferred to third parties under specific commercial agreements, they are anonymized through generalization \cite{0829/14/ENWP216}. In the following we provide the technical details to help the reader to deduce how much and to what extent these data are anonymized.

\paragraph{Privacy protection analysis}
The Route Service stores information about the routes (User Route) traveled by users in the Route Storage. This information is summarised and only made accessible to the user it belongs to. Its main purpose is to provide information concerning corresponding wallet charges and to let the user complaint if needed. Route Storage is designed to delete each record on a monthly basis, reflecting the \emph{data retention} policy enforced by GDPR.

On the other hand, the Completed Session Storage Service -- dedicated to each specific transport service -- stores the sessions (User Session) in the Completed Session Storage. The types of data included herein are listed in Table \ref{tab:completed_user_session_data}. As evident, they are anonymized through generalization relying on the \emph{age range} quasi-identifier. This technique produces a $k$-anonymous database. Specifically, $k$ may be derived relying on real statistics describing the public transport usage of a medium sized city.

\begin{table}[ht]
\centering
    \begin{tabular}{|l|r|r|}
    \hline
    Age range & Inhabitants & Bus users ($10.30\%$) \\
    \hline
    15-24 &	8$7247$ & $8986$ \\
    25-34 & $105143$ & $10830$ \\
    35-49 & $98989$ & $10196$ \\
    50-64 & $220554$ & $22717$ \\
    65-79 & $163279$ & $16818$ \\
    80+ & $85897$ & $8847$ \\
    \hline
    \end{tabular}
    \caption{Estimation of the number of citizens of Metropolitan City of Bologna that use bus service.}
    \label{tab:bologna_users}
\end{table}

\begin{table}[ht]
\centering
    \begin{tabular}{|l|r|}
    \hline
    & $k$ \\
    \hline
    Minimum (80+) & $8847$ \\
    Average & $13066$ \\
    Maximum (50-64) & $22717$ \\
    \hline
    \end{tabular}
    \caption{Maximum, minimum, and average $k$ values.}
    \label{tab:bologna_users-k-value}
\end{table}

For the purpose of this study, the analysis was carried out using data from the Metropolitan City of Bologna. Table \ref{tab:bologna_users} shows the number of inhabitants divided into age groups\footnote{https://www.tuttitalia.it/emilia-romagna/provincia-di-bologna/statistiche/popolazione-eta-sesso-stato-civile-2019/} and the estimated number of people using the bus service\footnote{https://statistica.regione.emilia-romagna.it/allegati/2020/dati-persone-che-utilizzano-mezzi-di-trasporto-istat-regioni-2019.csv}. Given the total number of sessions stored for a certain age range, we are thus able to derive the value of $k$. Considering as worst case the one where each person is coupled with only one session, $k$ would correspond to the number of people in a specific age group. Reflecting this principle, Table \ref{tab:bologna_users-k-value} shows the minimum, average and maximum value for $k$.

Concerning the Bologna case, a person is anonymous among $8847$ other, in the worst case. Moreover, if users were to make two ore more trips, these values would double, triple, and so forth.

\section{Conclusion}\label{sec:conclusion}
In this paper we introduced a novel ITS solution designed with privacy in mind and implementing the zero interaction paradigm. Following an exhaustive literature review we underlined the importance of integrated mobility systems within the smart city domain and we discussed the innovative aspects our system features. The mobility system has been effectively implemented and throughout the paper we detailed several technical aspects. To strengthen the discussion, UML paradigm was adopted to accompany the most relevant procedures. Both privacy preservation and the absence of human interaction were specifically discussed to highlight the differences with two prominent mobility systems operating within the European Union.
%
%
%
\bibliographystyle{splncs04}
\bibliography{biblio}

\end{document}